\documentclass[twocolumn,showpacs,preprintnumbers,amsmath,amssymb]{revtex4}

\usepackage{graphicx}
\usepackage{dcolumn}
\usepackage{bm}

\begin{document}

\preprint{Ver. 1.0}

\title{Optimal pair density functional for description of nuclei with large neutron excess}

\author{M. Yamagami,$^{1,2}$ Y. R. Shimizu,$^3$ and T. Nakatsukasa$^2$}
\affiliation{
$^1$Department of Computer Science and Engineering, University of Aizu,
Aizu-Wakamatsu, Fukushima 965-8580, Japan\\
$^2$ RIKEN Nishina Center, RIKEN,
Hirosawa 2-1, Wako, Saitama 351-0198, Japan\\
$^3$ Department of Physics, Kyushu University, Fukuoka 812-8581, Japan}


\date{\today}

\begin{abstract}
Toward a universal description of pairing properties in nuclei far from stability,
we extend the energy density functional by enriching the isovector density dependence 
in the particle-particle channel (pair density functional, pair-DF).
We emphasize the necessity of both the linear and quadratic isovector density terms. 
The parameters are optimized by the Hartree-Fock-Bogoliubov calculation 
for 156 nuclei of the mass number $A=118-196$ and 
the asymmetry parameter $(N-Z)/A<0.25$. 
We clarify that the pair-DF should include the isovector density dependence 
in order to take into account the effect of 
the isoscalar and isovector effective masses in the particle-hole channel consistently.
The different Skyrme forces can give the small difference in the pairing gaps 
toward the neutron drip line, if the optimal pair-DF consistent with 
the particle-hole channel is employed. 
\end{abstract}

\maketitle


\section{Introduction}

The energy density functional (EDF) theory provides a comprehensive
microscopic framework for description of 
bulk nuclear properties, low-lying excitations,
giant vibrations, and rotational excitations ~\cite{BH03}.
From the pioneering work by Vautherin and Brink \cite{VB72}, 
diverse endeavors have been made for finding the best EDF 
aiming at the description of the nuclear properties 
across the mass table.
For example, the Skyrme functional for the particle-hole (p-h) channel 
has been improved by taking into account 
the incompressibility modulus of nuclear matter \cite{KT80}, 
the spin- and spin-isospin channels \cite{GS81},
the deformation properties \cite{BQ82},
the spin-orbit terms~\cite{RF95}, and 
the isospin properties~\cite{CB98}.
Efforts to include the new terms such as
the tensor terms are also being made
(for the recent situation, see Ref.~\cite{LB07}).

The particle-particle (p-p) channel of the EDF 
(pair density functional, pair-DF) is also an indispensable element 
for description of nuclear systems \cite{DN96}.
The study of the nuclear matter predicts a very weak $^{1}S_0$
pairing at the normal density, and the pairing correlation
in finite nuclei is considered to be nuclear surface effects~\cite{BB05}.
The induced pairing interaction due to phonon exchange 
also enhances the surface effect \cite{BB05,TB02, BBG99}.
These facts suggest the density dependence of
the effective pairing force.

The standard parametrization of the pair-DF  has the isoscalar 
density ($\rho=\rho_n+\rho_p$) dependence only \cite{BH67,Ch76,KR79,BE91,TH95}.
The coupling constant should be constrained by the requirement to
reproduce the experimental data
such as masses, low-lying excited states, and rotational properties.
However, the functional form of the density dependence 
is still under discussion ~\cite{DN96,DN02}.

In nuclei near the $\beta$-stability line, the effect of the p-h field characterized by
the Fermi energy is much stronger than the p-p field. 
Therefore the pairing correlations can be treated 
within the BCS approximation~\cite{DF84,DN96}.
On the other hand, the strengths for the p-p and p-h channels 
become comparable in magnitude for weakly-bound nuclei
\cite{DN96,BD99,Ya05,DF84,MM05,Ma06,YS08}.
Therefore it is desirable to constrain the functional form of the pair-DF
by using the experimental data of unstable nuclei.

The isovector density ($\rho_1=\rho_n-\rho_p$) dependence 
can have sizable effects in nuclei apart from the $\beta$-stability line.
In Ref.~\cite{MS07}, the linear $\rho_1$ terms were introduced 
so as to simulate the neutron pairing gaps in symmetric 
and neutron matters obtained with either the bare interaction or the interaction
screened by the medium polarization effects. 
It was pointed out that 
the pairing properties in semi-magic nuclei
can be better described by the $\rho_1$-dependent pair-DF  
than that without $\rho_1$ terms \cite{MS08}.

We also recognized the importance 
of the linear $\rho_1$ term in the pair-DF \cite{YS08}.
By performing the Hartree-Fock-Bogoliubov (HFB) calculation 
with various coupling constants of the $\rho$ and $\rho_1$ terms,
we emphasized the strong sensitivity to the pairing properties
and the influence on rotational excitations 
in deformed nuclei near the neutron drip line.

In principle, it is desirable to derive the pair-DF from
the bare interaction based on the microscopic pairing theory 
including both the medium polarization effect 
and the surface phonon coupling effect in finite nuclei.
However, it seems very difficult at present 
in spite of recent progress toward this direction
~\cite{DH03,TB02,BBB05,PB08,Du04,DL08,FF08,GY08,GC08,CL06b}.

In this paper, we extend the pair-DF by including 
the linear and {\it quadratic} $\rho_1$ terms based on the phenomenological considerations.
The pair-DF is designed so as to reproduce the dependence of pairing gaps
on both the mass number $A$ and the asymmetry parameter $\alpha=(N-Z)/A$.
The parameters in the pair-DF are optimized so as to minimize the root-mean-square
(r.m.s.) deviation between the experimental and calculated pairing gaps.
The necessity of the $\rho_1$ dependence in pair-DF is emphasized
in connection with the the effective mass parameters.

This paper is organized as follows. In Sec.\ref{SEC-GLpair}, 
we briefly review the global properties of pairing gaps.
In Sec.\ref{SEC-MODEL}, our pair-DF including 
the linear and quadratic $\rho_1$ terms is introduced.
In Sec.~\ref{SEC-noRHO1}, we point out the drawback of
the pair-DF without the $\rho_1$ terms.
In Sec.~\ref{SEC-RHO1} and \ref{SEC-CSDF}, 
we investigate the role of the $\rho_1$ terms.
The parameters in the pair-DF are determined by 
the HFB calculation for 156 nuclei of  $A=118-196$ and $\alpha <0.25$. 
We clarify the close connection between the pair-DF and the effective masses
by the extensive analysis with 13 Skyrme parameters.
The choice of the pairing strength is discussed in Sec.~\ref{SEC-LAR}.
The conclusion is drawn in Sec.~\ref{SEC-CONC}.


\section{Global trend of pairing gaps} \label{SEC-GLpair}

\begin{figure}[t]
\center
\includegraphics[scale=0.65]{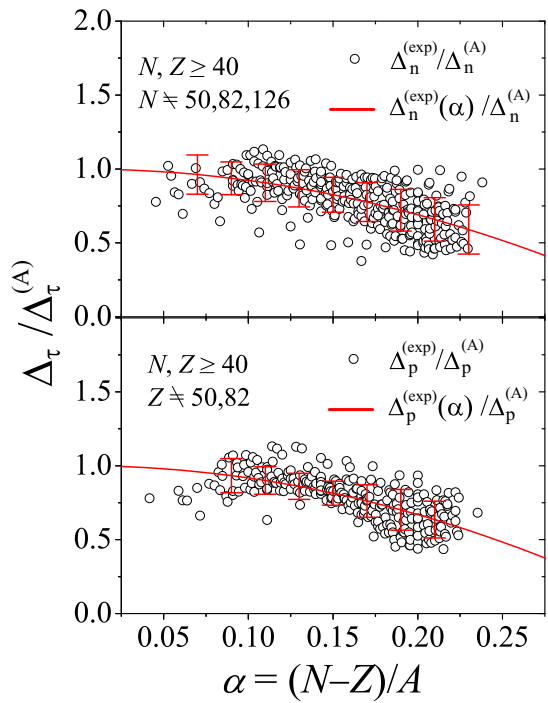}
  \caption{(Color online) Experimental neutron pairing gaps (upper panel) and 
  those of proton  (lower panel)
    in the region of $N, Z \ge 40$ 
   (except for nuclei with either $Z=50, 82$, or $N=50, 82, 126$)
    are shown  as a function of $\alpha$.
   The pairing gaps are divided by $\Delta_{\tau}^{(\text{A})}$. 
   The error bar represents the r.m.s. deviation 
   from the average trend $\Delta _{\tau}^{\text{(exp)}}(\alpha)$
   for each $\alpha$ with $\Delta \alpha=0.02$ interval. 
}
\label{fig-exp}
\end{figure}

We construct the pair-DF so as to reproduce 
the $A$- and $\alpha$-dependence of pairing gaps.
In Ref.\cite{VB84}, Vogel {\it et al}. pointed out that the 
experimental pairing gaps in the region of
$50<Z<82$ and $82<N<126$ can be well parametrized by 
$
\Delta \left( \alpha\right) 
= \left( {1 - 6.1\,\alpha ^2 } \right)\Delta^{(\text{A})}.   
\label{EQ-VOGEL}
$
Here the $A$ dependent part is given by
$\Delta^{(\text{A})}  = 7.2/A^{1/3}$ MeV.
This average $A$-and $\alpha$-dependence
holds for both neutron and proton pairing gaps.

We extend the analysis with up-to-date measured masses in
 the wider mass region of $N, Z \ge 40$ (except for
nuclei with either $Z=50, 82$ or $N=50, 82, 126$) \cite{AW03}.
The result is shown in Fig.~\ref{fig-exp}.
The average $A$- and $\alpha$-dependence is determined for
the neutron and proton pairing gaps separately by $\chi^2$-fitting;
\begin{eqnarray}
\Delta _{n}^{(\exp)} \left( \alpha \right) /\Delta _n^{(\text{A})} 
&\equiv &  C_{n,\text{exp}}^{(0)}  - C_{n,\text{exp}}^{(1)} \, \alpha^2 
\nonumber \\
&=&  {1 - 7.74\,\alpha ^2 } , 
 \label{EQ-DELN-EXP} \\
 \Delta _{p}^{(\exp)} \left( \alpha \right) / \Delta _p^{(\text{A})} 
&\equiv &  C_{p,\text{exp}}^{(0)}  - C_{p,\text{exp}}^{(1)} \, \alpha^2
\nonumber \\
&=&  {1 - 8.25\,\alpha ^2 },
  \label{EQ-DELP-EXP}
\end{eqnarray}
with $\Delta _n^{(\text{A})}  = 6.75/A^{1/3}$ MeV and 
$\Delta _p^{(\text{A})}  = 6.36/A^{1/3}$ MeV.
Here the experimental pairing gaps are extracted by
 the odd-even mass difference 
with the three-point staggering parameters \cite{DN02}.

The Coulomb force is an important building block of nuclear systems.
The 20 - 30 \% reduction of $\Delta_p$ by the self-consistent treatment 
of the Coulomb force was reported in Ref.~\cite{AE01}.  
The authors of Ref. \cite{LD09} also arrived at the same conclusion by  
performing the HFB calculation with the non-empirical pair-DF. 
On the other hand, the experimental evidence is unclear in our analysis.
The ratio is 
$\Delta_p^{(\text{exp})}(\alpha)/\Delta_n^{(\text{exp})}(\alpha) 
\approx 0.94 (1-0.51 \alpha^2 ) \ge 0.91$
for $0 \le \alpha \le 0.25$. 
This is smaller than the uncertainty of our analysis about 10 \%
shown by error bars in Fig.~\ref{fig-exp}.
The elaborate investigation is required to clarify the Coulomb effect.
Therefore we neglect this effect in our analysis
and leave it as an open problem in the future study.


\section{Model} \label{SEC-MODEL}

\subsection{Parametrization of pair-DF}

We extend the pair-DF
by including the linear and quadratic $\rho_1$ terms
in the following form;
\begin{eqnarray}
H_{\text{pair}} \left( \boldsymbol{r} \right) = \frac{{V_0 }}{4}\sum\limits_{\tau  = n,p} 
{g_\tau  \left[ {\rho ,\rho _1 } \right]
\left\{ {\tilde \rho _\tau  \left( \boldsymbol{r} \right)} \right\}^2 } 
\label{EQ-PLED}
\end{eqnarray}
with
\begin{eqnarray}
g_\tau  \left[ {\rho ,\rho _1 } \right] = 
1 - \eta_0 \frac{{\rho \left( \boldsymbol{r}  \right)}}{{\rho _0 }} 
- \eta _1 \frac{{\tau _3 \rho _1 \left( \boldsymbol{r}  \right)}}{{\rho _0 }} 
- \eta _2 \left( {\frac{{\rho _1 \left( \boldsymbol{r}  \right)}}{{\rho _0 }}} \right)^2. 
\label{EQ-pair-den-dep}
\end{eqnarray}
Here $\tau=n$ (neutron)  or $p$ (proton), and
$\rho_{0}=0.16$ fm$^{-3}$ is the saturation density of symmetric nuclear matter.
The $\tau_3 =$ $1$ (n) or $-1$ (p) in the linear $\rho_1$
term is introduced so as to preserves the charge symmetry of the pair-DF.
In nuclei with large $\alpha$, the $\rho_1$ terms
produce two effect for pairing correlations.
The one is the volume effect inside the nucleus,  which is relevant to all nuclei.
The other is the skin effect in nuclei apart from the $\beta$-stability line.

The pair-DF with $\eta_0=0.5$ and $\eta_1=\eta_2=0$ is one of 
the current standard parameterizations called  the mixed-type pairing force.
This pairing force reproduces the average $A$ dependence of pairing gaps \cite{DN02}.
We also justify this choice  in Sec.~\ref{SEC-RHO1}.
Therefore, we fix  $\eta_0=0.5$ unless otherwise noted.


\subsection{Setup}

We use the standard Skyrme interaction 
for the p-h channel in the HFB calculation.
The Skyrme SLy4 \cite{CB98} parametrization is mainly used.
In Sec.~\ref{SEC-CSDF}, we will extend our analysis with
13 Skyrme parameters. 

For the determination of $\eta_1$ and $\eta_2$, 
we perform the Skyrme-HFB calculation 
for 156 ground states of even-even, open-shell nuclei 
in the region of $Z=56-76$, and either $N=56-76$ or  $88-120$,
which covers the range of  $0<\alpha<0.25$.
We utilize the computer code of the Skyrme-HFB calculation 
developed by M. Stoitsov {\it et al}. \cite{SD05}.
Starting from the spherical, prolate and oblate initial conditions, 
the lowest energy solution is searched in the space of
axially symmetric quadrupole deformation.

We estimate the r.m.s. deviations between 
the experimental and calculated pairing gaps in order
to optimize the $\eta_1$ and  $\eta_2$.
The neutron and proton r.m.s. deviations are defined by
\begin{equation}
\sigma _\tau   = \left[ {\frac{1}{{N_{\tau}^{(\text{exp})} }}
\sum\limits_{\text{all data}} {\left( 
   \Delta _{\tau}- \Delta _{\tau}^{(\exp)}
\right)^2 } } \right]^{1/2}.
\end{equation}
The total r.m.s. deviation is also given by
\begin{eqnarray}
\sigma_{\text{tot}}=\left[\frac{N_{n}^{(\text{exp})}\sigma_{n}^2
+N_{p}^{(\text{exp})}\sigma_{p}^2}{N_{n}^{(\text{exp})}+N_{p}^{(\text{exp})} }\right]^{1/2}.
\end{eqnarray}
Here $N_{\tau}^{(\text{exp})}$ is the number of existing data of 
 $\Delta_{\tau}^{(\text{exp})}$ in the region of the present investigation; 
$N_{n}^{(\text{exp})}=93$ and $N_{p}^{(\text{exp})}=84$.
The theoretical pairing gap is defined by~\cite{BR00,YM01,Ma01}
\begin{eqnarray}
\Delta_{\tau} &=&-\int \mathrm{d}\boldsymbol{r}
\tilde{\rho}_{\tau}(\boldsymbol{r}) \tilde{h}_{\tau}(\boldsymbol{r})/\int \mathrm{d}\boldsymbol{r}
\tilde{\rho}_{\tau}(\boldsymbol{r}),
\end{eqnarray}
when the local pairing potential is given by
\begin{eqnarray}
\tilde{h}_{\tau}(\boldsymbol{r})=\frac{\partial}{\partial \tilde{\rho}_\tau (\boldsymbol{r})}
\int d \boldsymbol{r}' H_{\text{pair}} \left( \boldsymbol{r}' \right).
\end{eqnarray} 

We extract the coefficients $C_{\tau}^{(i)}$ which represent
the average $\alpha$-dependence of pairing gaps,
\begin{equation}
\Delta _{\tau} \left( \alpha \right) 
= \left(C_{\tau}^{(0)}  - C_{\tau}^{(1)}
\,\alpha ^2  \right)\Delta _{\tau}^{(\text{A})},
\end{equation}
by $\chi^2$-fitting analysis for $\Delta _{\tau}$ of the 156 nuclei. 
Here $\Delta _{\tau}^{(\text{A})}$ is the same quantity 
determined for Eqs.(\ref{EQ-DELN-EXP}) and (\ref{EQ-DELP-EXP}).

For each set of $(\eta_0,\eta_1,\eta_2)$,
the strength $V_0$ is fixed so as to reproduce
the $\Delta _{n}^{(\exp)}$ of $^{156}$Dy.
We use the abbreviation $V_0[\Delta_n(^{156}\text{Dy})]$
for this choice.
This nucleus has quadrupole deformation 
$\beta\approx0.28$ \cite{RP77}.
The experimental pairing gaps are
$\Delta_{n}^{(\text{exp})}=1.17$ MeV and 
$\Delta_{p}^{(\text{exp})}=0.98$ MeV, which are close to 
$\Delta_{n}^{(\text{exp})}(\alpha)=1.04$ MeV 
and $\Delta_{p}^{(\text{exp})}(\alpha)=0.96$ MeV
estimated by Eqs.(\ref{EQ-DELN-EXP}) and (\ref{EQ-DELP-EXP}).
The justification of $V_0$ will be discussed in Sec.~\ref{SEC-LAR}.

The cutoff quasiparticle energy $E_{\text{cut}}=50$ MeV is fixed in this paper.
We checked the dependence of $\sigma_\tau$ and $C_{\tau}^{(i)}$ 
on the $E_{\text{cut}}$ in Table~\ref{TAB-ED}.
The results with $E_{\text{cut}}=50$ and $75$ MeV 
agree within a few percent accuracy.
Here the parameters of the pair DF are fixed to be
$(\eta_0,\eta_1,\eta_2)=(0.5,0.2,2.5)$,
which are the optimal choice (See Sec.~\ref{SEC-RHO1}).

\begin{table}
\begin{center}
\begin{tabular}{c|c||ccc|cccc}  \hline \hline
$E_{\text{cut}}$ &  $V_0[\Delta_n(^{156}\text{Dy})]$ & $\sigma_{\text{tot}}$ & $\sigma_n$ 
& $\sigma_p$ & $C_n^{(0)}$ & $C_n^{(1)}$ & $C_p^{(0)}$ & $C_p^{(1)}$ \\
\hline \hline 
 50   & -346.5 & 0.17 & 0.16 & 0.17 & 1.08 & 9.42 & 1.00 & 8.13 \\
 75   & -320.0 & 0.17 & 0.16 & 0.18 & 1.07 & 9.26 & 1.01 & 8.44 \\
 \hline \hline
\end{tabular}
\end{center}
\caption{
The cutoff quasiparticle energy $E_{\text{cut}}$ dependence of 
the r.m.s. deviations [MeV] and the coefficients $C_{\tau}^{(i)}$ are listed.
The parameters $(\eta_0,\eta_1,\eta_2)=(0.5,0.2,2.5)$ are fixed. 
The strength $V_0$ [MeV fm$^{-3}$] is constrained by 
the $\Delta _{n}^{(\exp)}$ of $^{156}$Dy.
}
\label{TAB-ED}
\end{table}


\section{Pair-DF without $\rho_1$ term} \label{SEC-noRHO1}

\begin{figure}[t]
\center
\includegraphics[scale=0.65]{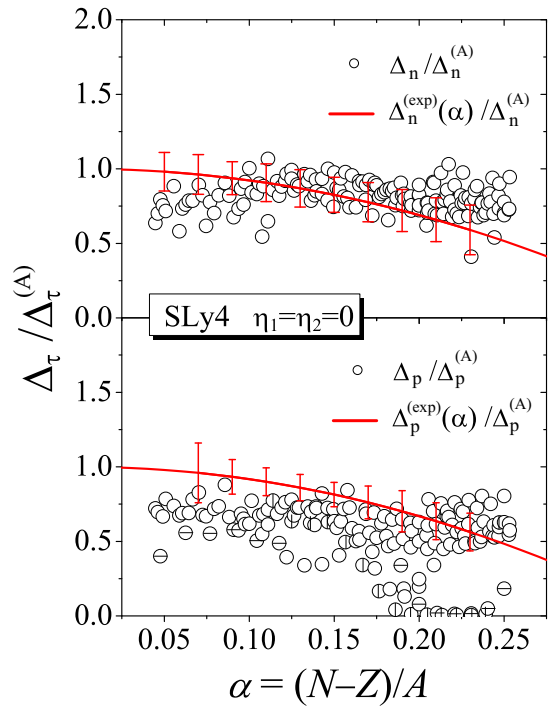}
  \caption{(Color online) Neutron and proton pairing gaps obtained with
  $\eta_0 =0.5$ and  $\eta_1=\eta_2=0$ are plotted as a function of $\alpha$.
 The pairing gaps are divided by $\Delta_{\tau}^{(\text{A})}$.
 The $_{60}$Nd and $_{70}$Yb isotopes
possessing the large proton shell gaps 
are indicated by circles with horizontal and vertical bars respectively 
in the bottom panel. 
}
\label{fig-DEL-MIX}
\end{figure}   

\begin{figure}[t]
\center
\includegraphics[scale=0.7]{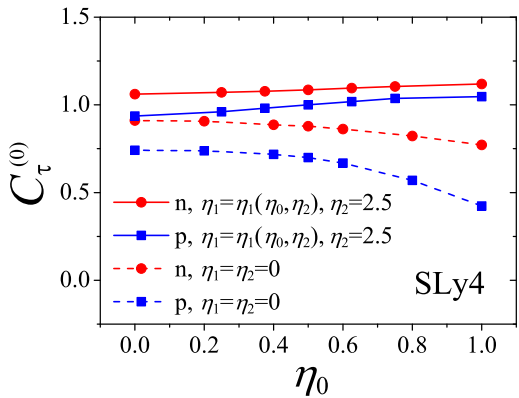}
  \caption{(Color online) 
 The coefficient $C^{(0)}_{\tau}$ obtained with
$\eta_1=\eta_2=0$ is shown as a function of $\eta_0$.
 The result with $\eta_2=2.5$ and $\eta_1(\eta_0,\eta_2)$ is compared.
Here $\eta_1(\eta_0,\eta_2)$ is the value of $\eta_1$
minimizing $\sigma_{\text{tot}}$ for each $(\eta_0,\eta_2)$
with $V_0[\Delta_n(^{156}\text{Dy})]$.
}
\label{fig-C0}
\end{figure}

\begin{figure}[t]
\center
\includegraphics[scale=0.7]{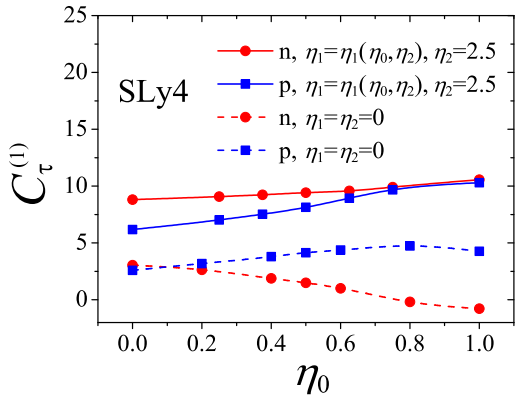}
  \caption{(Color online) The same with Fig.~\ref{fig-C0} but 
  for the coefficient $C^{(1)}_{\tau}$.}
\label{fig-C1}
\end{figure}   

\begin{figure}[t]
\center
\includegraphics[scale=0.7]{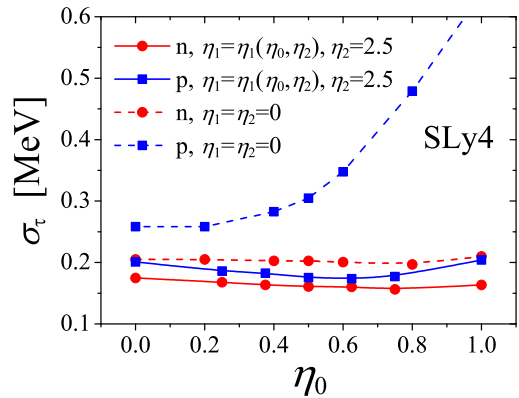}
  \caption{(Color online) 
The same with Fig.~\ref{fig-C0} but for $\sigma_\tau$.
}
\label{fig-sig-eta0}
\end{figure}

We show the drawback of the pair-DF without the $\rho_1$ terms.
The pairing gaps obtained with $\eta_0 =0.5$, $\eta_1=\eta_2=0$, 
and the strength $V_0[\Delta_n(^{156}\text{Dy})] = -324.0$ MeV fm$^{-3}$
are plotted in Fig.~\ref{fig-DEL-MIX}.
The extracted  $C_{\tau}^{(0)}$ and $C_{\tau}^{(1)}$ are shown 
by the dashed lines in Figs.~\ref{fig-C0} and \ref{fig-C1}.
The $\Delta_n$ and $\Delta_p$ are almost $\alpha$-independent.
The coefficient $C_{n}^{(1)}=1.11$ is much smaller 
than the experimental value $C_{n, \text{exp}}^{(1)}=7.74$.
Although the $C_{p}^{(1)}=3.74$ is larger than $C_{n}^{(1)}=1.11$,
this is due to the collapse of the pairing gap in weak pairing region.
Actually, it would be $C_{p}^{(1)}=1.38$, if we restrict the data to $\Delta_p > 0.25$ MeV.
Here the $_{60}$Nd and $_{70}$Yb isotopes
possessing the large proton shell gaps 
are indicated by circles with horizontal and vertical bars respectively 
in the bottom panel of Fig.~\ref{fig-DEL-MIX}.
The collapse of $\Delta_p$ is a drawback of the mean-field approximation \cite{RS80},
and can be overcome by performing the particle number projection (PNP).
The improvement for the $C_{\tau}^{(1)}$ values, however, 
can not be expected  by the PNP procedure \cite{SD03},
because the PNP method does not have any specific isovector effect.
Therefore, we neglect the effect of PNP in this study.

The $C_{n}^{(0)}=0.84$ is smaller than $C_{n, \text{exp}}^{(0)}=1$.
The $C_{p}^{(0)}=0.67$ is smaller than $C_{n}^{(0)}$ 
due to the quenching of $\Delta_p$ attributed to the neutron skin effect~\cite{YS08}: 
The neutron skin reduces the overlap between the form factor 
$[1-\eta_0 \rho(\boldsymbol{r})/\rho_0]$ and $\tilde{\rho}_p(\boldsymbol{r})$ 
in Eq.(\ref{EQ-PLED}).

The quenching of $\Delta_p$ due to the neutron skin effect 
becomes stronger with larger $\eta_0$ \cite{YS08}.
The $\sigma_\tau$ is shown as a function of $\eta_0$
by the dashed line in Fig.~\ref{fig-sig-eta0}.
Because the $\sigma_p$ rapidly increases with $\eta_0$,
the minimum of $\sigma_{\text{tot}}$ is absent.
In addition, the $C_\tau^{(0)}$ and $C_\tau^{(1)}$ 
remain small if restricted to $\eta_1=\eta_2=0$
(See Figs.~\ref{fig-C0} and \ref{fig-C1}).


\section{Role of $\rho_1$ dependence}  \label{SEC-RHO1}

\begin{figure}[t]
\center
\includegraphics[scale=0.7]{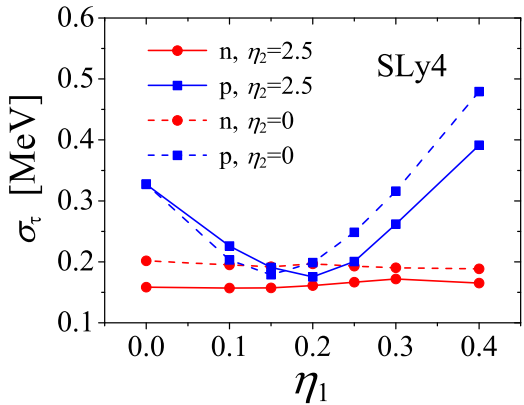}
\caption{(Color online) The r.m.s. deviations are shown as a function of $\eta_1$.
The results with either $\eta_2=0$ or $2.5$ are compared.}
\label{fig-sig-eta1}
\end{figure}

\begin{figure}[t]
\center
\includegraphics[scale=0.7]{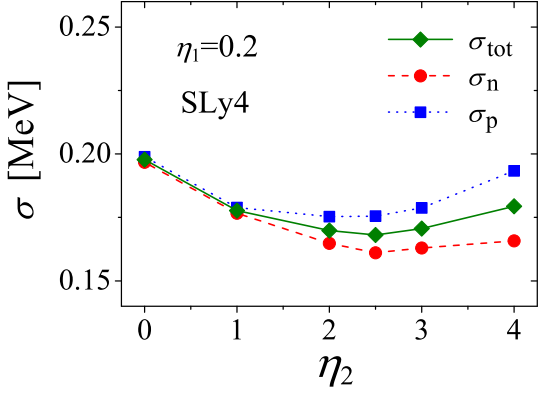}
\caption{(Color online) The r.m.s deviations at $\eta_1=0.2$ are shown
 as a function of $\eta_2$.}
\label{fig-05-02-x-sig}
\end{figure}

\begin{figure}[t]
\center
\includegraphics[scale=0.6]{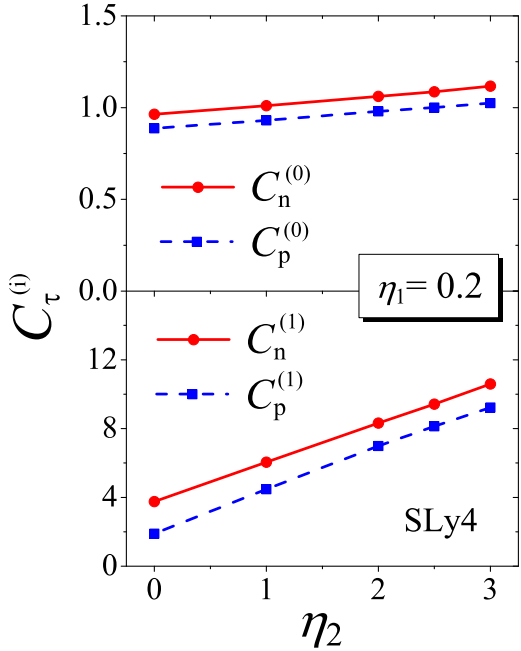}
\caption{(Color online) The coefficients $C_{\tau}^{(i)}$
at $\eta_1=0.2$ are shown as a function of $\eta_2$.}
\label{fig-DELav}
\end{figure}   

\begin{figure}[t]
\center
\includegraphics[scale=0.65]{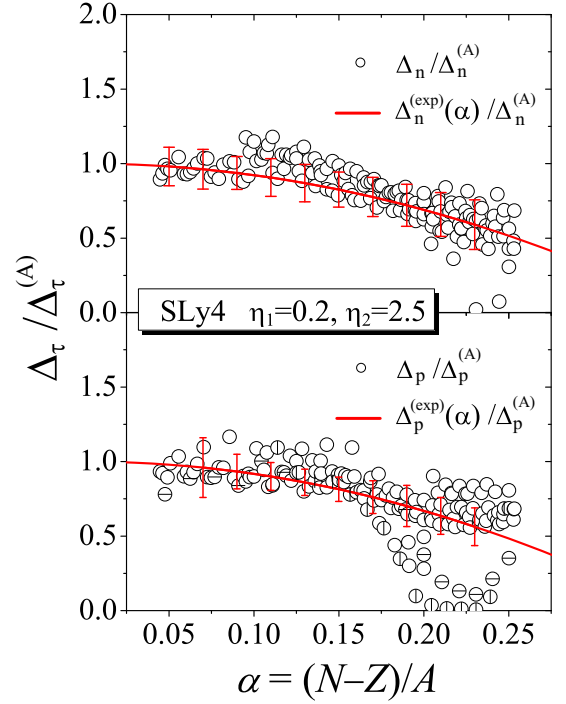}
  \caption{(Color online) The same as Fig.~\ref{fig-DEL-MIX}
  but with $(\eta_0,\eta_1,\eta_2)=(0.5,0.2,2.5)$.
}
\label{fig-DEL-OPT}
\end{figure}

\begin{table}
\begin{center}
\begin{tabular}{c|c||ccc|cccc}  \hline \hline
 criterion & $V_0$       &   $\sigma_{\text{tot}}$ & $\sigma_n$ & $\sigma_p$ 
& $C_n^{(0)}$ & $C_n^{(1)}$ & $C_p^{(0)}$ & $C_p^{(1)}$ \\
\hline \hline
 $V_{\text{opt}}(\text{def})$  & -344.0 &  0.16 &  0.14  & 0.18 & 1.03 & 9.71 & 0.99 & 7.64  \\
  $V_{\text{opt}}(\text{sph})$  & -308.0 &  0.50   & 0.47 & 0.52  & 0.59 & 6.23 & 0.52 &  4.62 \\
  \hline 
  $V_0[\Delta_n(^{156}\text{Dy})]$  &  -346.5 & 0.17 & 0.16 & 0.17 & 1.08 & 9.42 & 1.00 & 8.13 \\
 $V_0[\Delta_n(^{120}\text{Sn})]$   & -322.0 &   0.34  & 0.30 & 0.37  & 0.76 & 7.87 & 0.69 & 6.11  \\
        \hline \hline
Exp  & - & - & - & - & 1.00 & 7.74 & 1.00 & 8.25   \\
 \hline \hline
\end{tabular}
\end{center}
\caption{The r.m.s. deviations and the coefficients  $C_\tau^{(i)}$ 
obtained with the SLy4 force 
and the optimal parameters $(\eta_0,\eta_1,\eta_2)=(0.5,0.2,2.5)$ are shown.
The results with 
$V_{\text{opt}}(\text{def})$,
$V_{\text{opt}}(\text{sph})$, 
$V_0[\Delta_n(^{156}\text{Dy})]$,
and $V_0[\Delta_n(^{120}\text{Sn})]$
are compared.
The experimental values of $C_\tau^{(i)}$ are  also listed.
}
\label{TAB-V}
\end{table}

\begin{figure}[t]
\center
\includegraphics[scale=0.7]{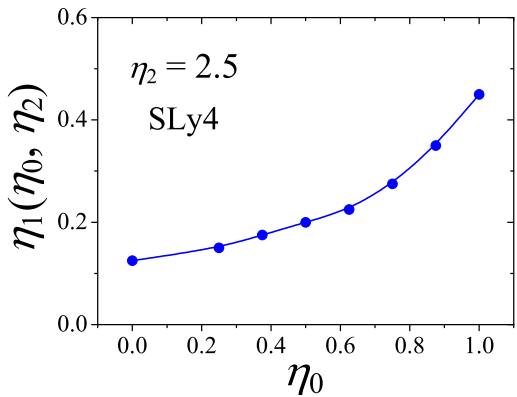}
\caption{(Color online) The value of $\eta_1$ minimizing the $\sigma_{\text{tot}}$
for each $(\eta_0,\eta_2)$ with $V_0[\Delta_n(^{156}\text{Dy})]$
is plotted as a function of  $\eta_0$.  The $\eta_2=2.5$ is fixed.
}
\label{fig-eta0-eta1}
\end{figure}

It is possible to compensate the quenching of $\Delta_p$
by using a stronger pairing strength for proton
(for example,  Ref.~\cite{GC09,BB09,BL09}).
However it violates the charge symmetry of the pair-DF.
This is the important symmetry  in the theoretical framework,
and indispensable for global description of pairing properties
 from neutron to proton drip line.

This consideration leads to introduction of the linear $\rho_1$ term
in Eq.~(\ref{EQ-pair-den-dep}).
This pair-DF preserves the charge symmetry.
The $\rho_1$ term induces the difference of 
the neutron and proton pairing strengths automatically \cite{YS08}.
The $\sigma_p$ has the minimum value at $\eta_1 = 0.15$, 
while the $\sigma_n$ is almost constant as a function of $\eta_1$.
This is shown by the dashed lines in Fig.~\ref{fig-sig-eta1}.
We obtain the $C_{n}^{(0)} = 0.93$ and  $C_{p}^{(0)} = 0.83$
with $\eta_1 =0.15$ and $\eta_2 = 0$, which are better than
those with $\eta_1 = \eta_2 = 0$.
However, the $C_{n}^{(1)} = 3.75$ and  $C_{p}^{(1)} = 1.89$
at $\eta_1 =0.15$ do not improve as a function of $\eta_1$.

The quadratic $\rho_1$ term in the pair-DF improves the r.m.s. deviations.
To see this, the r.m.s. deviations are plotted as a function of $\eta_1$
 while keeping $\eta_2=2.5$ in Fig.~\ref{fig-sig-eta1}.
Those with $\eta_1=0.2$ are plotted
as a function of $\eta_2$ in Fig.~\ref{fig-05-02-x-sig}.
The parameter set of $(\eta_1,\eta_2) = (0.2,2.5)$
simultaneously gives the minimum values of $\sigma_{\text{tot}}$, 
$\sigma_n$ and $\sigma_p$. 
They are smaller than those at $\eta_2=0$

The quadratic $\rho_1$ term also improves the $\alpha$-dependence. 
The   $C_{\tau}^{(0)}$ and  $C_{\tau}^{(1)}$ at $\eta_1=0.2$ are 
plotted as a function of $\eta_2$ in Fig.~\ref{fig-DELav}.
The  $C_{\tau}^{(0)}$ stays around 1.0,
while the $C_{\tau}^{(1)}$ increases linearly and reaches  
$C_{\tau}^{(1)} \approx C_{\tau, \text{exp}}^{(1)} \approx 8$
at $\eta_2 = 2.5$.

The pairing gaps obtained with
$(\eta_1,\eta_2) = (0.2,2.5)$ are shown in Fig.~\ref{fig-DEL-OPT}.
The r.m.s. deviations and the coefficients  $C_\tau^{(i)}$ are 
listed in Table \ref{TAB-V}.
We see the significant improvement compared to 
those with $\eta_1=\eta_2=0$. 

The optimized set of $(\eta_1,\eta_2)$ gives the justification
of the mixed type pairing force ($\eta_0=0.5$).
The $\sigma_\tau$ with $\eta_2 = 2.5$ and $\eta_1(\eta_0,\eta_2)$ 
is shown as a function of $\eta_0$ by the solid line in Fig.~\ref{fig-sig-eta0}.
The improvement over the choice $\eta_1=\eta_2=0$, 
especially the large reduction of $\sigma_p$, is obvious.
Therefore, the minimum of $\sigma_{\text{tot}}$ can appear at  $\eta_0 \approx 0.5$.
Here $\eta_1(\eta_0,\eta_2)$ is the value of $\eta_1$ minimizing 
$\sigma_{\text{tot}}$ for each $(\eta_0,\eta_2)$ with $V_0[\Delta_n(^{156}\text{Dy})]$.
The $\eta_1(\eta_0,\eta_2)$ at $\eta_2 = 2.5$  is shown as a function of $\eta_0$
in Fig.~\ref{fig-eta0-eta1}.

The coefficients  $C_{\tau}^{(0)}$ and  $C_{\tau}^{(1)}$
with $\eta_2 = 2.5$ and $\eta_1(\eta_0,\eta_2)$ 
are shown as a function of $\eta_0$ by the solid lines in Fig.~\ref{fig-C0} and \ref{fig-C1}.
The  $C_{\tau}^{(0)}$ is insensitive to $\eta_0$,
while the $C_{\tau}^{(1)}$ becomes close to the experimental value at $\eta_0 \approx 0.5$.


\section{Effective mass and $\rho_1$-dependence of pair-DF} \label{SEC-CSDF}

\subsection{Isoscalar and isovector effective masses}  
\label{subsec-emass}

\begin{table*}
\begin{center}
\begin{tabular}{l|ccc|cc||c|c||ccc|rr|rr}  \hline \hline
 Skyrme  & $m_v^* /m$  & $m_s^* /m$ & $\Delta m_1$ & 
  $W'_0/W_0$ & $\eta_{J}$ &  
  $\eta_1$ & $V_0[\Delta_n(^{156}\text{Dy})]$ & $\sigma_{\text{tot}}$ &
  $\sigma_{\text{n}}$ & $\sigma_{\text{p}}$ &  $C_n^{(0)}$ & $C_n^{(1)}$ & $C_p^{(0)}$ & 
$C_p^{(1)}$  \\ 
 \hline \hline
 SkM*  & 0.653 & 0.788 & -0.262 & 1 & 0 & 0.400 & -318.0 & 0.15 & 0.13 & 0.16 &  1.09 & 9.84 & 0.97 & 7.86 \\ \hline
 SGII    & 0.670 & 0.786 & -0.219 & 1 & 0  & 0.325 & -321.3 & 0.15 & 0.15 & 0.16 &  1.09 & 9.38 & 0.95 & 8.10 \\ \hline
 LNS  &  0.727 & 0.825 & -0.164 & 1 & 0 & 0.325 & -322.2 & 0.17 & 0.13 & 0.21 &  1.09 & 11.58 & 0.98 & 8.95 \\ \hline
 SkP   & 0.732 & 1.000 & -0.366 & 1 & 1  & 0.300 & -268.0 & 0.16 & 0.19 & 0.12 &  1.07 & 9.05 & 0.92 & 8.24 \\ \hline
 BSk17 & 0.780 & 0.798 & -0.028 & 1 & 1  & 0.200 & -313.5 & 0.14 & 0.14 & 0.13 &  1.08 & 8.68 & 0.94 & 6.94 \\ \hline
 SLy4  & 0.800 & 0.694 & 0.190 & 1 & 0  & 0.200 & -346.5 & 0.17 & 0.16 & 0.17 &  1.08 & 9.42 & 1.00 & 8.13 \\ \hline
 SLy5  & 0.800 & 0.697 & 0.184 & 1 & 1  & 0.200 & -342.0 & 0.16 & 0.15 & 0.16 &  1.06 & 8.94 & 0.98 & 8.38 \\ \hline
 SkI1  & 0.800 & 0.693 & 0.191 & 1 & 0  & 0.250 & -345.0 & 0.18 & 0.15 & 0.20 &  0.97 & 4.62 & 0.96 & 6.16 \\ \hline
 SkI4 & 0.800 & 0.649 & 0.290 & -0.985 & 0  & 0.275 & -364.5 & 0.18 & 0.16 & 0.20 &  1.02 & 8.25 & 0.97 & 7.57 \\ \hline
 SkO & 0.851 & 0.896 & -0.058 & -1.125 & 0  & 0.250 & -290.5 & 0.17 & 0.18 & 0.17 &  1.05 & 6.91 & 0.91 & 6.24 \\ \hline
 SkO'  & 0.871 & 0.896 & -0.032 & -0.576 & 1 & 0.200 & -289.0 & 0.16 & 0.16 & 0.15 &  1.03 & 6.62 & 0.89 & 5.64 \\ \hline
  \hline
  SkI3 & 0.800 & 0.574 &  0.493 & 0 & 0 & 0.125 & -397.4 & 0.20 & 0.20 & 0.20 &  1.04 & 7.26 & 0.98 & 8.63 \\ \hline
   SkT6 & 1.000 & 1.000 & 0.000 & 1 & 1  & 0.075 & -268.9 & 0.17 & 0.21 & 0.11 &  1.11 & 7.93 & 0.92 & 6.97 \\ \hline
 \hline
\end{tabular}
\end{center}
\caption{
The parameter set of the optimal pair-DF for each Skyrme parameterization is listed.
The optimal value of $\eta_1$ minimizing $\sigma_{\text{tot}}$
with the strength $V_0[\Delta_n(^{156}\text{Dy})]$ [MeV fm$^{-3}$],
the r.m.s. deviations [MeV], and the coefficients $C_{\tau}^{(i)}$ are shown. 
The parameters $(\eta_0,\eta_2)=(0.5,2.5)$ are fixed for them.
The effective masses $m_v^*$ and $m_s^*$
at the saturation density of symmetric nuclear matter,
the difference $\Delta m_{1} = m/m_s^* - m/m_v^*$, 
and the $W'_0/W_0$ and $\eta_{J}$ of the spin-orbit potential are also listed.
}
\label{TAB-Sk}
\end{table*}

\begin{figure}[t]
\center
\includegraphics[scale=0.7]{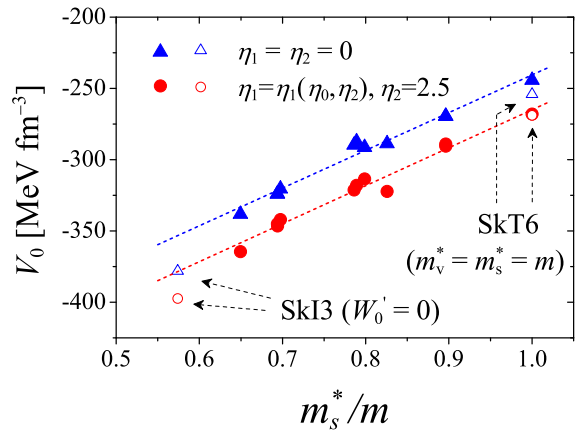}
\caption{(Color online) The strength $V_0$ reproducing the $\Delta_n^{(\text{exp})}$ of
$^{156}$Dy for each Skyrme force is plotted in relation to $m^*_s/m$. 
The results with $\eta_2=2.5$ and the optimal $\eta_1$ in 
Table \ref{TAB-Sk} are compared to those with $\eta_1=\eta_2=0$.
The $\eta_0=0.5$ is fixed.}
\label{fig-V-ms}
\end{figure}

\begin{figure}[t]
\center
\includegraphics[scale=0.7]{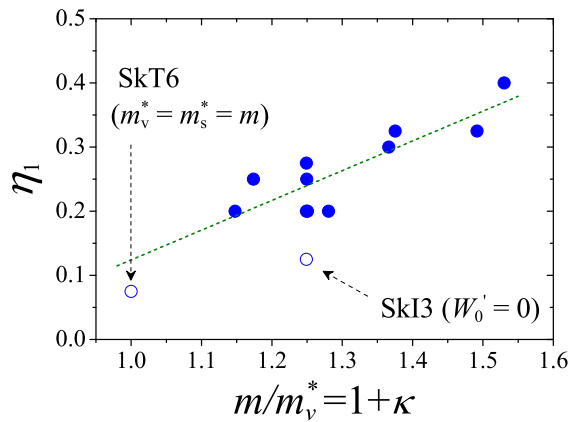}
\caption{(Color online) The optimal value of $\eta_1$ minimizing $\sigma_{\text{tot}}$ 
with $(\eta_0,\eta_2)=(0.5,2.5)$ and $V_0[\Delta_n(^{156}\text{Dy})]$ 
is shown in relation to $m/m_v^*$. See text for details.}
\label{fig-eta1-kappa}
\end{figure}

\begin{figure}[t]
\center
\includegraphics[scale=0.7]{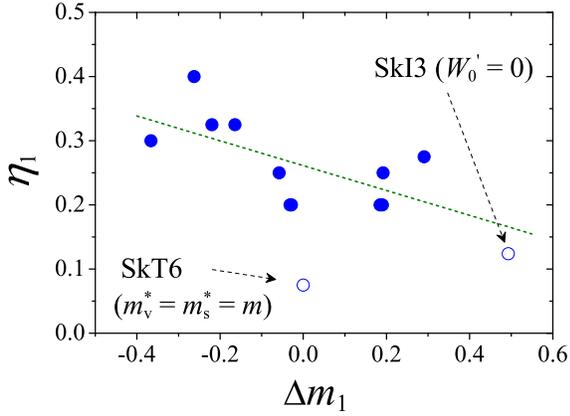}
\caption{(Color online) The same with Fig.~\ref{fig-eta1-kappa} but in relation to $\Delta m_1$.}
\label{fig-eta1-dm}
\end{figure}   

Pairing correlations are sensitive to the single-particle structure
around the Fermi level. 
For a suggestive example, the pairing gap is a function of $g G$
and given by $\Delta \propto e^{-1/gG}$ for $gG \ll 1$
in the schematic model of the seniority pairing force 
with the strength $G$ and the uniform single-particle level density $g$ \cite{RS80}.
On the other hand, the effective mass has a strong influence on the single-particle energies.
The average level density is proportional to the effective mass \cite{BB05}.
Therefore, we expect the close connection between the effective mass 
and the pair-DF  in order to reproduce the global trend of the experimental pairing gaps.

For the investigation, 
we extend our analysis with 13 Skyrme parameterizations; 
SkM* \cite{BQ82}, SGII \cite{GS81}, LNS \cite{CL06a}, 
SkP \cite{DF84}, BSk17 \cite{GC09}, SkT6 \cite{TB84}, 
SLy4, SLy5 \cite{CB98}, 
SkI1, SkI3, SkI4 \cite{RF95}, SkO, and SkO' \cite{RD99}.

The effective mass of the Skyrme force is given by
\begin{eqnarray}
\frac{\hbar^2}{2m_{\tau}^{*}({\bf r})}  
&=&  \frac{\hbar^2}{2m}+b_1 \rho - b'_1 \rho_\tau  \\
 &=& \frac{\hbar^2}{2m} \left\{ \frac{m}{m_{s}^{*}} 
+ \tau_3 I \Delta m_{1} \right\}
\end{eqnarray}
with  $I({\bf r}) =\rho_1/\rho$ and $\Delta m_{1}({\bf r})  = m/m_s^* - m/m_v^*$
\cite{BH03,LB06}.
The isoscalar and isovector effective masses  are defined by 
\begin{eqnarray}
\frac{m}{m_{s}^{*}({\bf r}) } &=& 1+\frac{2m}{\hbar^2} \left(b_1 - \frac{b'_1}{2} \right) \rho \\ 
\frac{m}{m_{v}^{*}({\bf r}) } &=& 1+\frac{2m}{\hbar^2} b_1 \rho = 1+\kappa.
\label{EQ-mv-kap}
\end{eqnarray}
The $m^*_v$ is directly connected to the enhancement factor $\kappa$ 
of the Thomas-Reiche-Kuhn sum rule \cite{BL79}.
We estimate the $m_s^*$ and $m_v^*$ at the saturation density
of the symmetric nuclear matter.
The  $m^*_v$ is a key parameter which determines 
the splitting of the neutron and proton effective masses
as a function of  $I$.
The parameters $b_1$ and $b'_1$ \cite{RD99} are given by
\begin{eqnarray}
b_1 &=& \frac{1}{8}\left[ t_1 \left(2+x_1 \right) 
+ t_2 \left(2+x_2 \right) \right]  \\
b'_1 &=& \frac{1}{8}\left[ t_1 \left(1+2 x_1 \right) 
- t_2 \left(1+2 x_2 \right) \right].
\end{eqnarray}

The spin-orbit potential ${\bf W}_\tau ({\bf r})$
also has the $\rho_1$ dependence.
The ${\bf W}_\tau ({\bf r})$ of Skyrme DF is defined by 
\begin{eqnarray}
{\bf W}_\tau ({\bf r}) &=&  \frac{1}{2} (W_0\nabla \rho +W'_0\nabla \rho_\tau )   
 +\eta_{J} {\bf W}_\tau^J  \nonumber \\
&=& \left(\frac{W_0}{2} + \frac{W'_0}{4}  \right) \nabla \rho 
+ \tau_3 \frac{W'_0}{4}  \nabla \rho_1   +\eta_{J} {\bf W}_\tau^J  \nonumber
\end{eqnarray}
where ${\bf W}_\tau^J  ({\bf r}) = C_0^J {\bf J} + \tau_3 C_1^J {\bf J}_1$
with the parameter $\eta_{J}$ of either 0 or 1.
The ${\bf J}$ (${\bf J}_1$) is the isoscalar (isovector) spin-current density. 
The $C_0^J$ and $C_1^J$ are given by 
$t_1$, $t_2$, $x_1$ and $x_2$ of the Skyrme parameter \cite{BD02}.
Most Skyrme functionals have the spin-orbit terms with $W_0=W'_0$.
However, the SkI4, SkO, and SkO' have the generalized $\rho_1$ dependence 
by introducing the different strengths $W_0$ and $W'_0$.
The SkI3 has $W'_0=0$.

We search the optimal value of $\eta_1$ which minimizes $\sigma_{\text{tot}}$
under the conditions; 1) the fixed $(\eta_0,\eta_2)=(0.5,2.5)$,  
and 2) the strength $V_0$ reproducing the $\Delta _{n}^{(\exp)}$ of $^{156}$Dy.
The numerical uncertainty is $\delta \eta_1 = 0.025$.
The results are summarized in Table \ref{TAB-Sk}.

The strengths $V_0$ reproducing the $\Delta _{n}^{(\exp)}$ of $^{156}$Dy 
are plotted in relation to  $m_s^*/m$ in Fig.~\ref{fig-V-ms}.
For $\eta_1 = \eta_2 = 0$, the $V_0$ increases linearly. 
This trend is in agree with the general consideration that
the pairing strength should be increased if the level density is low.
The relation is given by $V_0 = -505.05+264.47 m^*_s/m$ MeV fm$^{-3}$
with the correlation coefficient $r=0.99$, except for SkT6 and SkI3
(See Appendix~\ref{AP-CC} for the procedure of the correlation analysis). 
In general, the linear correlation disappears with the $\rho_1$ terms
due to the  $\eta_1$ and $\eta_2$ dependence of $V_0$.
However, it is interesting to mention that the linear correlation 
is recovered with  $\eta_2 = 2.5$ and the optimal $\eta_1$.
The extracted correlation is $V_0 = -531.45+266.46 m^*_s/m$
MeV fm$^{-3}$ with $r=0.99$.

The optimal values of $\eta_1$ are shown
in relation to $m/m_v^*$ in Fig.~\ref{fig-eta1-kappa}.
The linear correlation  between $\eta_1$ and $m/m_v^*$ is obvious,
irrespective of the choice of $W'_0$ and $\eta_J$.
The extracted relation is given by
\begin{eqnarray}
\eta_1 &=& -0.340+0.464 \, \frac{m}{m_v^*},
\label{EQ-ETA1}
\end{eqnarray}
or, in terms of the enhancement factor
\begin{eqnarray}
\eta_1 &=&  0.124 + 0.464 \, \kappa, 
\end{eqnarray}
except for SkI3 and SkT6.
The correlation coefficient between $\eta_1$ and $m/m_v^*$ is $r=0.85$.
This indicates that these parameters is almost linearly dependent.
The possible reason for the deviation of SkI3 and SkT6 is 
the special assumption on the Skyrme DF. 
The SkT6 sets $m_s^*=m_v^*=m$ by definition.
The SkI3 neglects the $\rho_1$ term in the spin-orbit potential by 
setting $W'_0=0$. 

During the optimization of the pair-DF, 
the $\eta_2=2.5$ for SLy4 is used for other Skyrme parameters
to avoid  the huge computational task.
However, the improvement obtained by the optimization 
of the parameter $\eta_2$ should be small. 
The effect can be estimated as follows:
We define the r.m.s. deviation of $C^{(1)}$ by
\begin{eqnarray}
\Delta C^{(1)} = \sqrt{\left<(C_\tau^{(1)}-C_{\tau, \text{exp}}^{(1)})^2 \right>}.
\end{eqnarray}
Here  $C_{\tau, \text{exp}}^{(1)}$ is the experimental value 
of Eqs.~(\ref{EQ-DELN-EXP}) and (\ref{EQ-DELP-EXP}).
The $\Delta C^{(1)}=1.5$
is obtained by taking the average  $<>$ over $\tau=n, p$ and  the 13 Skyrme parameters.
If the linearity $\Delta \eta_2 \approx \Delta C^{(1)} /2.3$ 
of Fig.~\ref{fig-DELav} and the parabolic approximation
for $\sigma_{\text{tot}}$ as a function of $\eta_2$ from Fig.~\ref{fig-05-02-x-sig}
are assumed for the other Skyrme parameters, 
the expected improvement for $\sigma_{\text{tot}}$ is 
about 0.002 MeV. 
In addition, we will show that the difference in the pairing gaps for different
Skyrme forces is small if the pair-DF with $\eta_2=2.5$ and 
the optimal $\eta_1$  is used in Sec.~\ref{SEC-LAR}.

From the present analysis, 
we conclude that the pair-DF should include the $\rho_1$ dependence 
in order to take into account the effect of the $m_s^*$ and $m_v^*$
for the global description of pairing correlations.

\subsection{$\Delta m_{1}$ dependence}

The effective masses $m_s^*$ and  $m_v^*$ have strong 
correlation with the $V_0$ and $\eta_1$ of the pair DF respectively.
On the other hand, 
the splitting of neutron and proton effective masses directly depends
on the local asymmetry parameter $I({\bf r})$ through 
the combination of $m_s^*$ and  $m_v^*$;
$\Delta m_{1} = m/m_s^* - m/m_v^*$.
If $\Delta m_1$ is negative, the $m_p^*$ ($m_n^*$)
is a decreasing (increasing) function of $I$.
In order to compensate the effect, the larger $\eta_1$
is necessary so as reproduce the same magnitude of
 the $\Delta_p^{(\text{exp})}(\alpha)$ and $\Delta_n^{(\text{exp})}(\alpha)$
(See Eqs.~(\ref{EQ-DELN-EXP}) and ~(\ref{EQ-DELP-EXP})).
On the other hand, the smaller $\eta_1$ is required
for the positive $\Delta m_1$.

This correlation can be seen in Fig.~\ref{fig-eta1-dm}.
The extracted correlation is
\begin{eqnarray}
\eta_1 = 0.261 - 0.193 \Delta m_1
\end{eqnarray}
with $r=-0.63$, except for SkT6 and SkI3.
Although this correlation is weaker than that 
between $\eta_1$ and $m/m_v^*$
due to the scattering of the $m_s^*/m$ value,
it is meaningful to conclude 
the linear dependence between  $\eta_1$ and $\Delta m_1$.
The $\eta_1 \approx 0.26$ at $\Delta m_1 \approx 0$ can be
the rough  estimation of the $\eta_1$ compensating 
the  artificial $\Delta_p$ suppression  due to the neutron skin effect.


\section{Choice of $V_0$}
\label{SEC-LAR}

\begin{figure}[t]
\center
\includegraphics[scale=0.7]{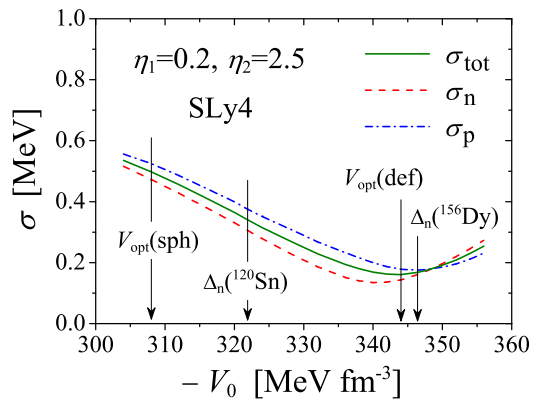}
\caption{(Color online) The r.m.s. deviations with SLy4 force and
 $(\eta_0,\eta_1,\eta_2)=(0.5,0.2,2.5)$ are shown as a function of $V_0$.
The strengths $V_{\text{opt}}(\text{def})$,
$V_{\text{opt}}(\text{sph})$, 
$V_0[\Delta_n(^{156}\text{Dy})]$,
and $V_0[\Delta_n(^{120}\text{Sn})]$ are indicated by arrows.}
\label{fig-V}
\end{figure}   

\begin{figure}[t]
\center
\includegraphics[scale=0.78]{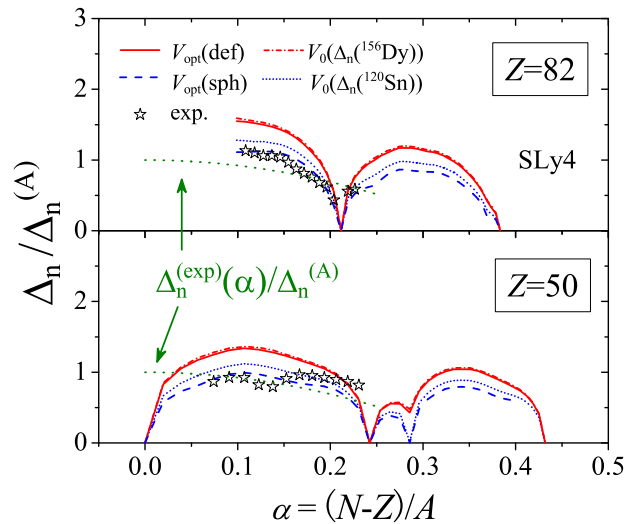}
\caption{(Color online) The neutron pairing gap of Sn and Pb isotopes 
with SLy4 force and $(\eta_0,\eta_1,\eta_2)=(0.5,0.2,2.5)$  are shown
as a function of $\alpha$. 
The results with the strengths; $V_{\text{opt}}(\text{def})$,  $V_{\text{opt}}(\text{sph})$, 
$V_{0}[\Delta_n(^{156}\text{Dy})]$, and $V_{0}[\Delta_n(^{120}\text{Sn})]$ are compared.
The experimental trend $\Delta_{n}^{(\text{exp})}(\alpha)$ is shown up to $\alpha < 0.25$
together with the experimental data. 
}
\label{fig-DELN-sph}
\end{figure}

\begin{figure}[t]
\center
\includegraphics[scale=0.78]{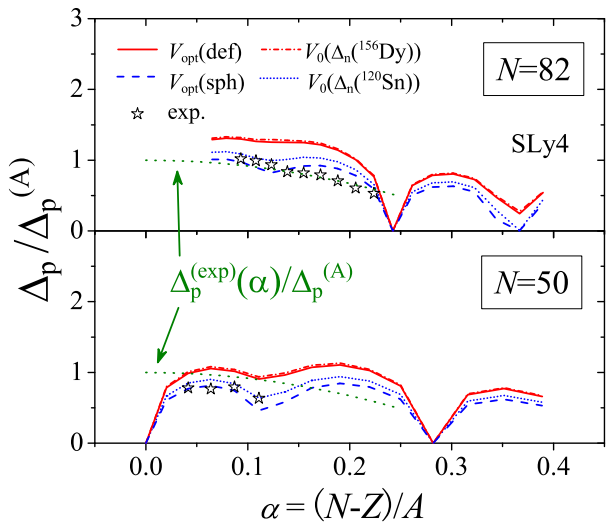}
\caption{(Color online) The same with Fig.~\ref{fig-DELN-sph}
but for the proton pairing gaps of $N=50$ and $82$ isotones.
}
\label{fig-DELP-sph}
\end{figure}

\begin{figure}[t]
\center
\includegraphics[scale=0.78]{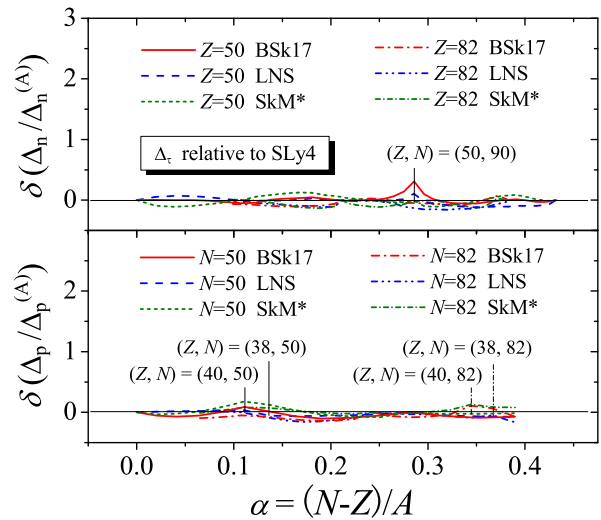}
\caption{(Color online) 
The difference in the pairing gaps for the different Skyrme parameters.
See text for details.
}
\label{fig-dDEL}
\end{figure}

It is desirable to optimize the strength $V_0$ 
for experimental data in wide region of nuclear chart. 
However, the procedure demands heavy computational efforts.
Therefore the $V_0$ is usually fixed so as to reproduce 
a pairing gap of specific nucleus.
Several authors adopted the $\Delta_n^{(\text{exp})}$ of $^{120}$Sn 
\cite{SD00,SD03,BO05}.
The strength
$V_0[\Delta_n(^{120}\text{Sn})] = -322.0$ MeV fm$^{-3}$
with  SLy4 force and $(\eta_1, \eta_2)=(0.2, 2.5)$ gives
$\sigma_{\text{tot}}=0.34$ MeV (see Table.~\ref{TAB-V}).
On the other hand, 
if the $V_0$ is fixed in deformed region, for example,  
$V_0[\Delta_n(^{156}\text{Dy})] = -346.5$ MeV fm$^{-3}$
the r.m.s. deviation reduces to $\sigma_{\text{tot}}=0.17$ MeV.
The $V_0$ dependence of the $\sigma_{\text{tot}}$ is
shown in Fig.~\ref{fig-V}.
The optimal value $V_{\text{opt}}(\text{def})=-344.0$ MeV fm$^{-3}$
for the deformed nuclei is close to the $V_0[\Delta_n(^{156}\text{Dy})]$.

In Figs.~\ref{fig-DELN-sph} and \ref{fig-DELP-sph},
the $\Delta_n$ of Sn and Pb isotopes and 
the $\Delta_p$ of $N=50$ and $82$ isotones 
obtained with  SLy4 force and $(\eta_1, \eta_2)=(0.2, 2.5)$  are shown.
The results with 
$V_{\text{opt}}(\text{def})$,
$V_0[\Delta_n(^{156}\text{Dy})]$,
$V_0[\Delta_n(^{120}\text{Sn})]$ are compared.
The difference of the results with $V_{\text{opt}}(\text{def})$ and
$V_0[\Delta_n(^{156}\text{Dy})]$ is negligible along the isotopic and isotonic chains.
However, the choice of $V_0[\Delta_n(^{156}\text{Dy})]$ overestimates
the experimental pairing gaps \cite{AW03} in these spherical nuclei.
The strength optimized only for the spherical nuclei is 
$V_{\text{opt}}(\text{sph})=-308.0$  MeV fm$^{-3}$.
This is 10.4 \% weaker than $V_{\text{opt}}(\text{def})$.

It is an open problem to construct 
the pair-DF which allows us to describe the pairing properties
along the chains of semi-magic nuclei
at the same quality achieved for deformed region
\cite{BB09,BL09,DB02a}.
The authors of Ref.~\cite{BB09} considered that
the overestimation in spherical nuclei may be partly attributed 
to the effect of the particle number fluctuation.
They showed that the HFB calculation with the approximate
particle number projection using the Lipkin-Nogami method
improves the agreement with experiment for spherical nuclei.
We do not discuss this point further in detail, and
the choice of $V_0[\Delta_n(^{156}\text{Dy})]$ is employed in this work.

The strengths $V_0$ for other Skyrme parameters
are also determined by the same procedure.
The $\sigma_{\text{tot}}$ with the choice of $V_0[\Delta_n(^{156}\text{Dy})]$
is almost the same quality compared to SLy4 (see Table \ref{TAB-Sk}).
In Fig.~\ref{fig-dDEL}, the $\Delta_n /\Delta_n^{(\text{A})}$  
of Sn and Pb isotopes and the $\Delta_p /\Delta_p^{(\text{A})}$ of $N=50$ and $82$ 
isotones obtained with Skyrme BSk17, LNS and SkM* are shown.
The $\eta_1$ and  $V_0[\Delta_n(^{156}\text{Dy})]$ in Table \ref{TAB-Sk}
are used for the Skyrme forces. 
For comparison, the value obtained with the SLy4 force is subtracted;
$\delta (\Delta_\tau /\Delta_\tau^{(\text{A})})(X)
=\Delta_\tau /\Delta_\tau^{(\text{A})}(X) - \Delta_\tau /\Delta_\tau^{(\text{A})} (\text{SLy4})$
for $X=$ BSk17, LNS and SkM*.
Their $(\Delta m_1, m_v^*/m)$ are $(0.190,0.800)$, $(-0.028,0.780)$, 
$(-0.164,0.727)$, and $(-0.262,0.653)$ for SLy4, BSk17, LNS, and SkM* respectively.
The $\sigma_{\text{tot}}$ of the BSk17 is smallest,
and the $\kappa$ of SkM* is largest  in this work.
The LNS parametrization was built to match the $I$ dependence of the effective 
masses and the neutron matter EOS predicted by Br\"uckner-Hartree-Fock  calculation. 

In spite of the variety of $\Delta m_1$ and $m_v^*$, 
the $\delta (\Delta_\tau /\Delta_\tau^{(\text{A})})$ is small
along the isotopic and isotonic chains, 
except for around the subshell closure;
$N=90$ for $\Delta_n$, and  $Z=38$ and $40$ for $\Delta_p$.
This is because the pairing correlations are sensitive to
the single-particle structure around the subshell closure. 

The small $\delta(\Delta_n /\Delta_n^{(\text{A})})$ can be expected 
due to the weak sensitivity of $\sigma_n$ to $\eta_1$ if the strength $V_0$
is constrained by $\Delta_{n}^{(\text{exp})}$ of a specific nucleus \cite{YS08}.
This is seen in Fig.~\ref{fig-sig-eta1}.
Authors of Ref.~\cite{LB06} also pointed out that
the $\Delta_n$ of Sn and Pb isotope chains are insensitive to $\Delta m_1$
by performing the HFB calculation with the mixed type pairing force
and various Skyrme forces. 
On the other hand, the fine tuning of $\eta_1$ is indispensable 
for the small  $\delta(\Delta_p /\Delta_p^{(\text{A})})$  due to the sensitivity of $\sigma_p$
to $\eta_1$. This is shown in Fig.~\ref{fig-sig-eta1},
and discussed in Ref.~\cite{YS08}.

The $\eta_2=2.5$ for SLy4 is commonly used for other Skyrme parameters
(see Sec.~\ref{SEC-CSDF}). This is an approximation in our analysis.
However, the small $\delta (\Delta_\tau /\Delta_\tau^{(\text{A})})$
as a function $\alpha$ means that the difference in the $\alpha$ 
dependence of the pairing gaps due to the different Skyrme forces
 can be small with the fixed $\eta_2=2.5$.

We refer the pair-DF with the parameters in Table \ref{TAB-Sk}
as the optimal one for each Skyrme parameterization.
The optimal pair-DF with $\rho_1$ dependence is
constructed aiming at unique description of pairing properties toward the neutron drip line.
Our prescription is based on the phenomenological considerations.
In this sense, our conclusion is tentative.
However, the optimal pair-DF can preserves the good descriptive power 
of the neutron excess dependence of pairing correlations,
and provide the certain foundation for the further improvement 
with experimental data of  nuclei with larger neutron excess.

\section{Conclusion}   \label{SEC-CONC}

We proposed a new pair-DF by introducing the $\rho_1$ dependence. 
We emphasized the necessity of both the linear and quadratic $\rho_1$ terms 
in the pair-DF for the global description of pairing correlations; 
namely the dependence on both the mass number $A$ 
and the neutron excess $\alpha=(N-Z)/A$.

To optimize the parameters in the pair-DF,
we performed the HFB calculation for 156 nuclei of  $A=118-196$ and $\alpha<0.25$. 
By the extensive investigation with 13 Skyrme parameterizations, 
we clarify that the pair-DF should include the $\rho_1$ dependence 
in order to take into account
the effect of the $m_s^*$ and $m_v^*$ in the p-h channel:
The $\eta_1$ and $m/m_v^*$ is linearly dependent, and
the pairing strength $V_0$ linearly increases as a function of $m_s^*/m$ 
with the optimal set of  $(\eta_0,\eta_1,\eta_2)$.
The relationship between the optimal $\eta_1$ and 
the splitting of the neutron and proton effective masses is also discussed.
The  $V_0$ is fixed so as reproduce the $\Delta_n^{(\text{exp})}$ of $^{156}$Dy.
With this choice, we can obtain the almost minimum value 
of the total r.m.s. deviation between the experimental and calculated pairing gaps. 
The different Skyrme forces with the optimal pair-DF can give
the small difference in the pairing gaps toward the neutron drip line.

In this paper, we concentrated on the analysis of pairing gaps
in finite nuclei based on the phenomenological consideration.
To obtain the deeper insight to the $\rho_1$ terms in the pair-DF, 
it is interesting to investigate pairing correlations 
in asymmetric nuclear matter by comparing 
the up-to-date calculations with 3-body force
and correlations beyond the mean-field approximation.
This analysis is a future subject.

\begin{acknowledgments}
This work is supported in part by the JSPS Core-to-Core Program,
International Research Network for Exotic Femto Systems (EFES)
 and by Grant-in-Aid for Scientific Research on Innovative
Areas (No. 20105003) and by the Grant-in-Aid for Scientific Research (B)
(No. 21340073).
We are grateful to H. Sagawa, K. Yabana, and K. Matsuyanagi 
for valuable discussions.
The numerical calculations were carried out on Altix3700 BX2
and SX8 at YITP in Kyoto University and the RIKEN Super Combined 
Cluster (RSCC) in RIKEN.
\end{acknowledgments}

\appendix

\section{Correlation analysis} 
\label{AP-CC}

We introduce the correlation coefficient $r$ for
a data set $(x,y)=\{(x_i, y_i)\}$ $(i=1,2,\cdots,n)$.
It is defined by 
\begin{eqnarray} r=\frac{\sum_{i=1}^n (x_i-\bar{x})(y_i-\bar{y})}
{\sqrt{\sum_{i=1}^n (x_i-\bar{x})^2} \sqrt{\sum_{i=1}^n (y_i-\bar{y})^2} }.
\end{eqnarray}
The coefficient $r$ can take a real value of $-1 \le r \le 1$.
In the limit of $r=1$ or $-1$, the data set is linearly dependent.
On the other hand, the correlation between $x$ and $y$ is weak
if $r$ is close to zero.

\end{document}